\documentclass[twocolumn,aps,prl,preprintnumbers,superscriptaddress,amsmath,amssymb,showpacs]{revtex4}

\usepackage{graphicx}
\usepackage{dcolumn}
\usepackage{bm}

\newcommand{\etal}{{\it et al.}}

\def\Journal#1#2#3#4{{#1} {\bf #2}, #3 (#4)}

\def\NIMA{Nucl.\ Instrum.\ Methods A}
\def\PLB{Phys.\ Lett.\ B}
\def\PRL{Phys.\ Rev.\ Lett.}
\def\PRC{Phys.\ Rev.\ C}
\def\PRD{Phys.\ Rev.\ D}

\def\AAA{Astron.\ Astrophys.}
\def\APJ{Astrophys.\ J.}
\def\APP{Astropart.\ Phys.}

\begin{document}

\title{Search for Cosmic-Ray Antideuterons}

\author{H.\thinspace Fuke}
\email[E-mail address: ]{fuke@balloon.isas.jaxa.jp}
\affiliation{Institute of Space and Astronautical Science (ISAS/JAXA), Sagamihara, Kanagawa 229-8510, Japan}
\author{T.\thinspace Maeno}
\altaffiliation[Present address: ]{BNL, Upton, NY 11973}
\affiliation{Kobe University, Kobe, Hyogo 657-8501, Japan}
\author{K.\thinspace Abe}
\altaffiliation[Present address: ]{ICRR, Kashiwa, Chiba 227-8582, Japan}
\affiliation{Kobe University, Kobe, Hyogo 657-8501, Japan}
\author{S.\thinspace Haino}
\affiliation{High Energy Accelerator Research Organization (KEK), Tsukuba, Ibaraki 305-0801, Japan}
\author{Y.\thinspace Makida}
\affiliation{High Energy Accelerator Research Organization (KEK), Tsukuba, Ibaraki 305-0801, Japan}
\author{S.\thinspace Matsuda}
\affiliation{The University of Tokyo, Tokyo 113-0033, Japan}
\author{H.\thinspace Matsumoto}
\affiliation{The University of Tokyo, Tokyo 113-0033, Japan}
\author{J.\thinspace W.\thinspace Mitchell}
\affiliation{NASA, Goddard Space Flight Center, Greenbelt, MD 20771}
\author{A.\thinspace A.\thinspace Moiseev}
\affiliation{NASA, Goddard Space Flight Center, Greenbelt, MD 20771}
\author{J.\thinspace Nishimura}
\affiliation{The University of Tokyo, Tokyo 113-0033, Japan}
\author{M.\thinspace Nozaki}
\affiliation{Kobe University, Kobe, Hyogo 657-8501, Japan}
\author{S.\thinspace Orito}
\altaffiliation{Deceased}
\affiliation{The University of Tokyo, Tokyo 113-0033, Japan}
\author{J.\thinspace F.\thinspace Ormes}
\altaffiliation[Present address: ]{Univ.\thinspace of Denver, Denver, CO 80208}
\affiliation{NASA, Goddard Space Flight Center, Greenbelt, MD 20771}
\author{M.\thinspace Sasaki}
\affiliation{NASA, Goddard Space Flight Center, Greenbelt, MD 20771}
\author{E.\thinspace S.\thinspace Seo}
\affiliation{University of Maryland, College Park, MD 20742}
\author{Y.\thinspace Shikaze}
\altaffiliation[Present address: ]{JAERI, Tokai, Ibaraki 391-1195, Japan}
\affiliation{Kobe University, Kobe, Hyogo 657-8501, Japan}
\author{R.\thinspace E.\thinspace Streitmatter}
\affiliation{NASA, Goddard Space Flight Center, Greenbelt, MD 20771}
\author{J.\thinspace Suzuki}
\affiliation{High Energy Accelerator Research Organization (KEK), Tsukuba, Ibaraki 305-0801, Japan}
\author{K.\thinspace Tanaka}
\affiliation{High Energy Accelerator Research Organization (KEK), Tsukuba, Ibaraki 305-0801, Japan}
\author{K.\thinspace Tanizaki}
\affiliation{Kobe University, Kobe, Hyogo 657-8501, Japan}
\author{T.\thinspace Yamagami}
\affiliation{Institute of Space and Astronautical Science (ISAS/JAXA), Sagamihara, Kanagawa 229-8510, Japan}
\author{A.\thinspace Yamamoto}
\affiliation{High Energy Accelerator Research Organization (KEK), Tsukuba, Ibaraki 305-0801, Japan}
\author{Y.\thinspace Yamamoto}
\altaffiliation[Present address: ]{KEK,\thinspace Tsukuba,\thinspace Ibaraki\thinspace 305-0801,\thinspace Japan}
\affiliation{The University of Tokyo, Tokyo 113-0033, Japan}
\author{K.\thinspace Yamato}
\affiliation{Kobe University, Kobe, Hyogo 657-8501, Japan}
\author{T.\thinspace Yoshida}
\affiliation{High Energy Accelerator Research Organization (KEK), Tsukuba, Ibaraki 305-0801, Japan}
\author{K.\thinspace Yoshimura}
\affiliation{High Energy Accelerator Research Organization (KEK), Tsukuba, Ibaraki 305-0801, Japan}
\date{\today}

\begin{abstract}
We performed a search for cosmic-ray antideuterons 
using data collected during four BESS balloon flights from 1997 to 2000. 
No candidate was found. 
We derived, for the first time, an upper limit of 
1.9$\times 10^{-4}$~(m$^2$s~sr~GeV/nucleon)$^{-1}$ 
for the differential flux of cosmic-ray antideuterons, 
at the 95\% confidence level, 
between 0.17 and 1.15~GeV/nucleon at the top of the atmosphere. 
\end{abstract}


\maketitle

The possible presence of various species of antimatter 
in the cosmic radiation 
can provide evidence of sources and processes 
important for both astrophysics and elementary particle physics. 

For example, discovery of a single antihelium nucleus 
in the cosmic radiation would offer clear evidence 
for a baryon symmetric cosmology. 
Despite extensive and ongoing searches, 
none has ever been found~\cite{bib:Sasaki,bib:AMSHebar}. 

Similarly, the spectral form and magnitude of the antiproton ($\bar{p}$) 
spectrum could provide evidence for a number of possible primary sources, 
including evaporating primordial black holes 
(PBHs)~\cite{bib:MakiPBAR,bib:BarrauPBAR} 
and annihilating neutralino dark matter~\cite{bib:Bergstrom,bib:DonatoSUSY} 
as well as a baryon symmetric cosmology. 
However, recent results from 
the BESS experiment~\cite{bib:OritoBESS97,bib:MaenoBESS98,bib:AsaokaBESS00} 
imply that most of the $\bar{p}$'s in the cosmic radiation 
are not from primary sources, 
but rather are secondary products of the energetic collisions 
of Galactic cosmic rays with the interstellar medium. 
The data do suggest that below $\sim$1~GeV there is an excess of 
$\bar{p}$'s above expectation from a purely secondary origin, 
but the situation is far from clear. 
Model calculations of the secondary spectrum still have 
ambiguities~\cite{bib:MakiPBAR,bib:Bergstrom,bib:Bieber,bib:Moskalenko,bib:DonatoSec} 
and statistical errors of the currently measured low-energy $\bar{p}$ 
spectrum are not small enough to provide clarity. 
The accuracy of both calculations and measurements 
needs substantial improvement. 

While antideuterons ($\bar{d}$'s) have never been detected 
in the cosmic radiation, they can be produced by the same sources 
as $\bar{p}$'s and may be of both secondary or primary origin, 
with the latter providing evidence for sources such as 
PBHs and annihilating neutralino dark matter. 
The low energy range below $\sim$1~GeV/nucleon offers a unique window 
in the search for cosmic-ray primary $\bar{d}$'s 
because it has a greatly reduced background from secondary 
$\bar{d}$'s~\cite{bib:Chardonnet,bib:Donato,bib:GAPS}, 
as compared with secondary $\bar{p}$'s. 
Thus, the unambiguous detection of a single $\bar{d}$ 
below $\sim$1~GeV/n would strongly suggest the existence of 
novel primary origins. 
Hence, cosmic-ray $\bar{d}$'s have an advantage over cosmic-ray 
$\bar{p}$'s as a probe to search for primary origins. 

In this paper we report on a search for $\bar{d}$'s 
carried out with four balloon flights of the BESS instrument 
from 1997 to 2000. 
Using data from these flights, we report for the first time 
an upper limit on the differential flux of cosmic-ray $\bar{d}$'s 
and discuss this result in the context of expectation 
from evaporating PBHs. 

The BESS detector was 
designed~\cite{bib:OritoAstromag,bib:AkiraPropose} 
and developed~\cite{bib:Ajima} 
as a high-resolution spectrometer 
with the large geometrical acceptance 
and strong particle-identification capability 
required for antimatter searches. 
A uniform magnetic field of 1~Tesla is generated 
by a thin superconducting solenoid. 
The field region is filled with tracking detectors 
consisting of a jet-type drift chamber (JET) 
and two inner drift chambers (IDCs). 
Tracking is performed by fitting up to 28 hit points 
in these drift chambers, 
resulting in a rigidity ($R$) resolution of 0.5\% at 1~GV. 
The upper and lower time-of-flight scintillator hodoscopes 
(TOFs) 
measure the velocity ($\beta$) and the energy loss (d$E$/d$x$). 
The time resolution of each counter is 55~ps, 
which yields a 1/$\beta$ resolution of 1.4\%. 
A threshold-type \v{C}erenkov counter 
with a silica-aerogel radiator ($n$=1.03 in 1997 and $n$=1.02 thereafter) 
can reject $e^{-}/\mu^{-}$ events, 
which can be backgrounds for the detection of $\bar{p}$'s and $\bar{d}$'s, 
by a factor of $\sim$10$^3$. 

Four balloon flights were carried out 
in northern Canada, 1997 through 2000: 
from Lynn Lake to Peace River 
where the geomagnetic cutoff rigidity ranges from 0.3 to 0.5~GV. 
Data for the $\bar{d}$ search were taken for 
live time of 15.8, 16.8, 27.4, and 28.7 hours 
in 1997, 1998, 1999, and 2000 respectively, 
at altitudes around 36~km, 
corresponding to $\sim$5~g/cm$^2$ in residual atmospheric depth. 
The data acquisition sequence was initiated by a first-level trigger, 
which is generated by a coincidence between hits of 
the top and bottom TOFs with a threshold set at 1/3 
of the pulse height from minimum ionising particles. 
In addition to biased trigger modes~\cite{bib:Ajima,bib:MaenoBESS98} 
enriching negatively charged particles, 
one of every 60 (30 in 2000) first-level triggered events 
were recorded as unbiased samples. 

The concept of the off-line analysis 
is similar to that used for the $\bar{p}$ selection 
described in Ref~\cite{bib:MaenoBESS98}. 
For events of both negative and positive curvature, 
same selections were applied 
to detect clear $\bar{d}$ and deuteron ($d$) candidates. 
The selected $d$'s were used to estimate selection efficiencies 
for $\bar{d}$'s. 
At the first step, 
we selected events with a single downward-going, passing-through track 
which is fully contained inside the fiducial volume 
with restricted number of TOF hits, 
in order to reject interacted events as well as albedo particles. 
At the second step, 
in order to eliminate backgrounds such as large-angle scattered events 
by ensuring good quality of $R$ and $\beta$ measurements, 
we applied several cuts on tracking and timing 
measurement quality parameters such as: 
(i) the number of used hits 
and the reduced $\chi^2$ of the trajectory fitting, 
and (ii) the consistency between the JET track, hits in the IDCs, 
and the TOF timing information. 

\begin{figure}[t]
  \begin{center}
    \includegraphics[width=85mm]{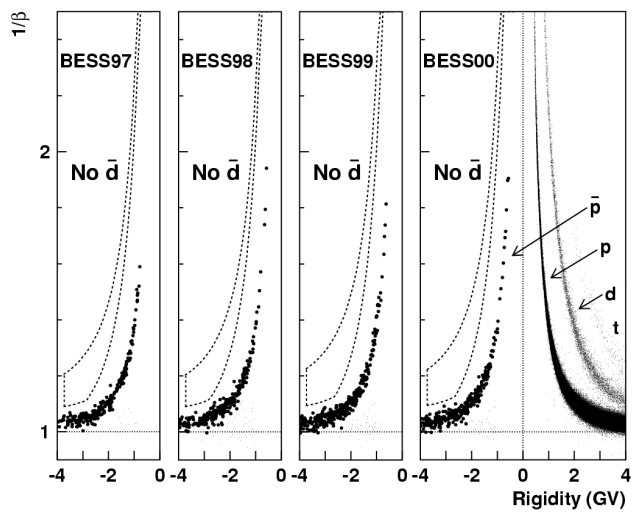}
  \end{center}
  \caption{
    The surviving single-charge events in the data of each flight. 
    The dotted curves define the $\bar{d}$ mass bands. 
    The dotted vertical lines at $\sim$-3.7~GV corresponds to 
    $E_{2}$ shown in Fig.~\ref{fig:sensitivity}. 
  }
  \label{fig:beta2d98}
\end{figure}

In order to identify $\bar{d}$'s, 
d$E$/d$x$ measurements 
inside the TOFs and the JET were required 
to be consistent with $d$'s as a function of $R$. 
In addition, the \v{C}erenkov veto was applied to 
reduce the $e^{-}/\mu^{-}$ background contamination. 
Thereafter, the mass of the incident particle was reconstructed 
using the measured $\beta$ and $R$. 
Figure~\ref{fig:beta2d98} shows the 1/$\beta$ vs.\ $R$ plots 
of the events which survived all the above selections 
(for 1997 -- 1999, only the negative rigidity events are shown). 
The $\bar{d}$ selection region was determined 
by the mirror position of the $d$ band, 
which was defined to have a uniform selection efficiency of 99\%. 
In order to avoid the contamination (or misidentification) of $\bar{p}$'s, 
the region overlapped by the $\bar{p}$ band was excluded. 
The exclusion band was defined so that 
it had a uniform selection efficiency for $\bar{p}$'s and 
the possible $\bar{p}$ contamination from the whole data set 
was just one event or less 
in an estimation using the 1/$\beta$ distributions of 
positively charged events. 
The contamination of other negatively charged particles 
(mainly $e^{-}/\mu^{-}$) was estimated 
to be less than 0.1 event. 
The possibility of the spill over of positively charged particles 
into the negative side is negligible in the considered rigidity region. 
In Fig.~\ref{fig:beta2d98} (panel BESS00), 
candidates of protons ($p$'s) and $\bar{p}$'s 
selected by the same procedure 
are superimposed. 

As shown in Fig.~\ref{fig:beta2d98}, 
clear mass bands of $p$'s, $d$'s, tritium and $\bar{p}$'s can be seen. 
However, no $\bar{d}$ candidate exists 
within the expected selection region. 

Since no $\bar{d}$ candidate was found, 
we calculated the resultant upper limit 
on the $\bar{d}$ flux~\cite{bib:commentlimit}, 
$\Phi_{\bar{d}}$,
which is given by:
$
\Phi_{\bar{d}}=
N_{\rm obs}
/
\big|S\Omega
~\varepsilon_{\rm total}~(1-\delta_{\rm sys})\big|_{\rm min}
/
T_{\rm live}
/
(E_{2}-E_{1})
$. 
The live time, $T_{\rm live}$, was directly measured by 
a 1~MHz-clock pulse generator and scalers throughout the flights. 
As the number of the observed $\bar{d}$ events, $N_{\rm obs}$, 
we took 3.09 for the calculation of the 95\% C.L. 
upper limit. 
We did not consider the effect of the possible background contamination 
($\alt$1 event), because the background estimation 
still has an ambiguity since it is difficult to evaluate the amount of 
the tail distribution strictly. 
$E_{1}$ and $E_{2}$ denote the energy range of the limit 
at the top of the atmosphere (TOA). 
The energy measured at the instrument was traced back to the one 
at the TOA by correcting the ionisation energy loss. 
In order to obtain the most conservative limit, 
the minimum value of 
$(S\Omega~\varepsilon_{\rm total}~(1-\delta_{\rm sys}))$ 
was used, 
where $S\Omega$ is the geometrical acceptance, 
$\varepsilon_{\rm total}$ is the total detection efficiency, 
and $\delta_{\rm sys}$ is the total systematic uncertainty. 
The $\varepsilon_{\rm total}$ can be written as 
$
\varepsilon_{\rm total}=
\varepsilon_{\rm trig}
~\varepsilon_{\rm 1}
~\varepsilon_{\rm 2}
~\varepsilon_{\rm pid}
~\varepsilon_{\rm acc}
~\varepsilon_{\rm air}
$~. 
The efficiency of the first step selection 
including the effects of inelastic interactions in the instrument 
($\varepsilon_{\rm 1}$), 
the survival probability in the residual atmosphere 
($\varepsilon_{\rm air}$), 
and the $S\Omega$ 
were calculated by the BESS Monte Carlo (MC) based on GEANT/GHEISHA. 
Since there are no experimental data of $\bar{d}$ interactions in material, 
we incorporated the $\bar{d}$ in the code 
under the following assumptions: 
(i)~The inelastic cross sections of $\bar{d}$ can be estimated 
by scaling those of $\bar{p}$ using an empirical model 
of hard spheres with overlaps~\cite{bib:HS-Bradt,bib:HS-Grigalashvili}, 
which is described as: 
$\sigma(A_i,A_t)\propto
[A_i^{1/3}+A_t^{1/3}-0.71\times(A_i^{-1/3}+A_t^{-1/3})]^2$; 
where $\sigma(A_i,A_t)$ is the cross section of an incident particle 
with atomic weight $A_i$ to a target with atomic weight $A_t$. 
(ii)~When an inelastic interaction occurs, 
$\bar{d}$ is always fragmented or annihilated. 
(iii)~Other effects of energy loss, multiple scattering, 
bremsstrahlung and $\delta$-rays 
are described as are those of other nuclei. 
This hard sphere model is known to reproduce data on nuclear interactions 
for various combinations of $A_i$ and $A_t$ 
including light nuclei such as 
$p$/$d$~\cite{bib:HS-Grigalashvili,bib:HS-Abdrahmanov} 
and antinuclei including 
$\bar{p}$/$\bar{d}$~\cite{bib:HS-Hoang,bib:RHICdbar}. 
We adopted this model to estimate the $\sigma(\bar{d}, A_t)$ 
scaling from the $\sigma(\bar{p}, A_t)$ 
described in Ref.~\cite{bib:AsaokaBeam}. 
The efficiency of the the second step selection ($\varepsilon_{\rm 2}$) 
was estimated by using both 
the unbiased data and the BESS MC. 
The trigger efficiency ($\varepsilon_{\rm trig}$) 
was obtained by using the unbiased data 
and a detector beam-test data~\cite{bib:AsaokaBeam}. 
The efficiency of particle identification 
($\varepsilon_{\rm pid}$) 
was estimated using 
the unbiased $d$ samples of each flight 
under the assumption that the $\bar{d}$ candidate should 
behave similarly to $d$ 
except for deflection in the symmetrical configuration of BESS. 
Typical values at 0.5~GeV/n are: 
$\varepsilon_{\rm trig} \sim 90$\%, 
$\varepsilon_{\rm 1} \sim 60$\%, 
$\varepsilon_{\rm 2} \sim 70$\%, 
$\varepsilon_{\rm pid} \sim 98$\%, 
$\varepsilon_{\rm air} \sim 85$\%, 
and $S\Omega \sim $ 0.25~m$^2$sr. 
The probability of events without 
any hits or tracks by another accidental incident particle, 
$\varepsilon_{\rm acc}$, 
was derived to be $\sim$94\% 
by samples taken by the random trigger 
which was issued at once a second throughout the flights. 

\begin{figure}[t]
  \begin{center}
  \includegraphics[height=45mm]{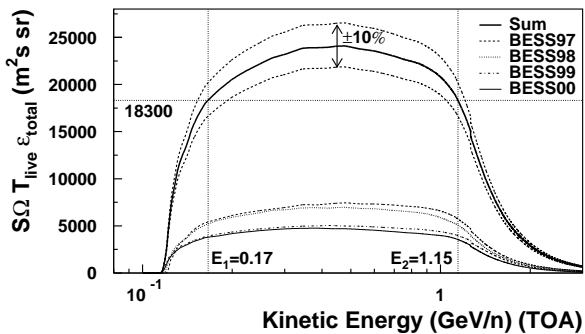}
  \caption
      {Effective exposure factors of each flight data and their sum. The systematic uncertainty is shown as a $\pm$10\% width band. $E_{1}$ and $E_{2}$ denote the energy range of the limit. 
      }
  \label{fig:sensitivity}
  \end{center}
\end{figure}

Figure~\ref{fig:sensitivity} shows the calculated effective exposure factor. 
The decrease of the factor at the low-energy side 
is mainly caused by the decrease of the geometrical acceptance, 
the decrease of mean free path through the detector, 
and the increase of large-angle scattering. 
The major reason for the decrease at the high-energy side 
is the decrease of $\varepsilon_{\rm pid}$ 
due to the overlap of 1/$\beta$ distributions 
between $\bar{d}$'s and $\bar{p}$'s. 
The combined systematic uncertainty, 
which was estimated to be $\delta_{\rm sys}$~$\sim$~10\% 
with less energy dependence, 
is also shown in the figure. 
Dominant systematic uncertainties were the uncertainties 
in the evaluation of $\varepsilon_{\rm 1}$, 
$\varepsilon_{\rm 2}$ and $\varepsilon_{\rm air}$, 
all of which were discussed using the MC simulation. 
The energy range of $E_{1}$ -- $E_{2}$ 
was chosen to be 
0.17 -- 1.15~GeV/n, 
where the exposure factor is highest 
and has relatively little energy dependence. 

\begin{figure}[t]
  \begin{center}
    \includegraphics[height=110mm]{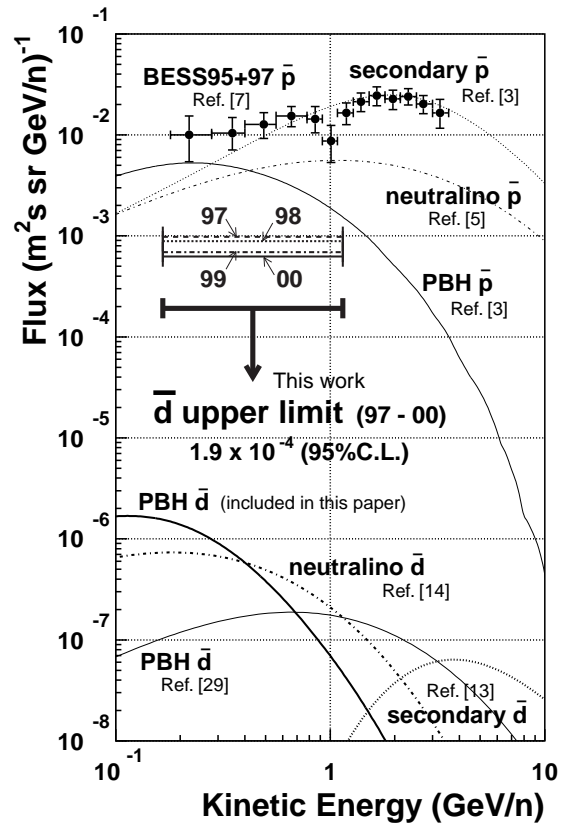}
  \end{center}
  \caption{
    Obtained upper limit on the $\bar{d}$ flux in comparison with 
    the PBH-$\bar{d}$ spectrum calculated in this paper. 
    A $\bar{p}$ spectrum measured by BESS and some theoretically predicted 
    spectra of $\bar{p}$ and $\bar{d}$ are shown for reference. 
  }
  \label{fig:dblimit}
\end{figure}

The resultant upper limit $\Phi_{\bar{d}}$ 
for 1997, 1998, 1999, 2000, and 
the integrated flight data 
were calculated to be 
9.8, 8.9, 6.9, 6.2, and 1.9 $\times 10^{-4}$~(m$^2$s~sr$^2$GeV/n)$^{-1}$ 
respectively (Fig.~\ref{fig:dblimit}). 
These are the most conservative limits 
with no assumptions on the $\bar{d}$ spectrum shape. 
If we assume a uniform $\bar{d}$ energy spectrum 
and use a mean inverse exposure factor, 
the summed upper limit was evaluated to be 
1.6$\times 10^{-4}$ in the same energy range, 
and 1.4$\times 10^{-4}$ in the range 0.13 -- 1.44~GeV/n 
where the upper limit is minimized under this assumption. 
Since our detection efficiency of $\bar{d}$ 
is less dependent on the energy, 
the upper limit is less dependent on 
the assumption of the $\bar{d}$ spectrum shape. 
In the following discussions, 
we use the most conservative one (1.9$\times 10^{-4}$). 

As described in Ref.~\cite{bib:MakiPBAR}, 
only PBHs that are close to explosion 
and exist within a few kpc of the solar system 
can contribute to the observed flux. 
Therefore, the $\bar{d}$ upper limit leads directly to the upper limit 
on the explosion rate of local PBHs, ${\mathcal R}_{\rm PBH}$. 
In order to obtain ${\mathcal R}_{\rm PBH}$ 
from the $\bar{d}$ upper limit, 
we have calculated an expected PBH-$\bar{d}$ spectrum 
through the following steps: 
(i) the emission rate of particles from PBHs, 
(ii) the fragmentation rate to form $\bar{d}$'s, 
(iii) the source spectrum, 
(iv) the propagation process, 
and (v) the effect of the solar modulation. 
The calculations except for (ii) were based on the calculation of 
PBH-$\bar{p}$ spectrum 
described in Ref.~\cite{bib:MakiPBAR}. 
The step (ii) was performed by using the frequently-used 
``coalescence model'' (e.g.\ Ref.~\cite{bib:Butler-Yazaki-Nagle}). 
According to this model, the production probability of $\bar{d}$'s 
in momentum space, ${\rm d}^3 n_{\bar{d}}/{\rm d}p^3$, 
can be expressed as the product of 
those of $\bar{p}$'s and antineutrons: 
$\gamma~\frac{{\rm d}^3 n_{\bar{d}}}{{\rm d}p^3}=
\frac{4}{3}\pi p_0^3
~(\gamma~\frac{{\rm d}^3 n_{\bar{p}}}{{\rm d}p^3}) 
~(\gamma~\frac{{\rm d}^3 n_{\bar{n}}}{{\rm d}p^3}) 
~\approx 
~\frac{4}{3}\pi p_0^3
~(\gamma~\frac{{\rm d}^3 n_{\bar{p}}}{{\rm d}p^3})^2
$, 
where $p_0$ is the ``coalescence momentum'' 
which must be determined from experiments. 
We assumed $p_0=130$~MeV/$c$ from the data of $\bar{d}$ production 
in $e^+$/$e^-$ annihilation~\cite{bib:ARGUS}. 
The solar modulation in step (v) was estimated 
by using the numerical solution of the spherically symmetric model 
proposed by Fisk~\cite{bib:Fisk}. 
The solar modulation parameter, $\phi$, was determined 
to fit the $p$ spectrum measured in the same BESS flights as: 
500, 610, 648, and 1334~MV in 1997, 1998, 1999, and 2000 
respectively~\cite{bib:MaenoBESS98,bib:AsaokaBESS00}. 
The calculated PBH-$\bar{d}$ flux for 1997 
on the assumption of 
${\mathcal R}_{\rm PBH}=2.2\times 10^{-3}$~pc$^{-3}$yr$^{-1}$ 
is shown in Fig.~\ref{fig:dblimit}. 
The difference between our PBH-$\bar{d}$ flux and 
the one in Ref.~\cite{bib:BarrauDBAR} 
mainly comes from the different assumptions on the propagation model, 
similar to the case of the difference in the PBH-$\bar{p}$ spectrum 
between Ref.~\cite{bib:MakiPBAR} and Ref.~\cite{bib:BarrauPBAR} 
that is described in Ref.~\cite{bib:BarrauPBAR}. 

Here, we place 
the upper limit on the ${\mathcal R}_{\rm PBH}$ 
to be $1.8\times 10^0$~pc$^{-3}$yr$^{-1}$, 
which is five orders of magnitude more stringent than the sensitivity 
for 50-TeV $\gamma$-ray bursts~\cite{bib:Tibet}. 
The limit on ${\mathcal R}_{\rm PBH}$ leads to an upper limit on 
the density parameter of PBHs in the Universe, $\Omega_{\rm PBH}$, 
to be 1.2$\times 10^{-6}$. 
The initial mass spectrum of PBHs was assumed to have 
a $-\frac{5}{2}$ power-law form, 
and the PBH spatial distribution was assumed to be proportional to 
the mass density distribution of dark matter 
within the galactic halo~\cite{bib:MakiPBAR}.

As a conclusion, 
we have searched for cosmic-ray $\bar{d}$'s 
with the BESS flight data obtained between 1997 and 2000. 
No $\bar{d}$ candidate has been detected. 
We placed, for the first time, an upper limit 
of $1.9\times 10^{-4}$~(m$^2$s~sr~GeV/n)$^{-1}$ (95\% C.L.) 
on the differential flux of cosmic-ray $\bar{d}$'s 
in an energy range of 0.17 -- 1.15~GeV/n at the top of the atmosphere. 
In consequence, we derived 
an upper limit of $1.8\times 10^{0}$ pc$^{-3}$yr$^{-1}$ (95\% C.L.) 
on the explosion rate of local PBHs 
and an upper limit of $1.2\times 10^{-6}$ (95\% C.L.) 
on the density parameter of PBHs. 

These upper limits regarding PBHs are 
two orders of magnitude looser than 
those derived from the $\bar{p}$ flux~\cite{bib:MakiPBAR}. 
However, further sensitive searches could push down 
the limits from $\bar{d}$'s below the ones from $\bar{p}$'s, 
because the low-energy range has a greatly reduced 
background from secondary $\bar{d}$'s. 
Astrophysical consequences of our $\bar{d}$ search will motivate 
further sensitive searches for $\bar{d}$'s 
as well as further advances in the physics of primary origins, 
in connection with cosmology and elementary particle physics. 

\begin{acknowledgments}
We would like to thank NASA, NSBF, KEK, ISAS 
and ICEPP for their continuous support. 
This experiment was supported 
by a Grants-in-Aid, 
KAKENHI (9304033, 11440085, and 11694104), 
from MEXT and by Heiwa Nakajima Foundation in Japan; 
and by NASA SR\&T research grants in the USA. 
\end{acknowledgments}

\bibliography{test}

\begin{thebibliography}{}
\bibitem{bib:commentlimit} The upper limit 
  on the absolute $\bar{d}$ flux is the best indicator 
  of the sensitivity of the search for the primary sources 
  (rather than the $d$/$\bar{d}$ flux ratio for example).
\end{thebibliography}

\end{document}